\documentclass[%
 aip,
 amsmath,amssymb,
 preprint,%
]{revtex4-1}
\newcounter{algctr}
\renewcommand{\thealgctr}{\Alph{section}.\arabic{algctr}} 
\newcommand{\algcaption}[1]{%
  \refstepcounter{algctr}%
  \par\medskip\noindent\textbf{Algorithm~\thealgctr: #1}\par\smallskip%
}

\usepackage{graphicx}
\usepackage{xcolor}
\usepackage{float}
\usepackage{caption}
\usepackage{subcaption}
\usepackage{booktabs}

\usepackage{algpseudocode}
\usepackage[colorinlistoftodos]{todonotes}

\draft 

\makeatletter
\@booleanfalse\titlepage@sw
\def\frontmatter@abstract@produce{%
  \par
  \addvspace{\frontmatter@preabstractspace}%
  \begingroup
   \dimen@\baselineskip
   \setbox\z@\vtop{\unvcopy\absbox}%
   \advance\dimen@-\ht\z@\advance\dimen@-\prevdepth
   \@ifdim{\dimen@>\z@}{\vskip\dimen@}{}%
  \endgroup
  \begingroup
   \prep@absbox
   \unvbox\absbox
   \post@absbox
  \endgroup
  \@ifx{\@empty\mini@notes}{}{\mini@notes\par}%
  \addvspace\frontmatter@postabstractspace
}%
\makeatother

\begin{document}


\title{Reinforcement-learning control of turbulence transition in the modified Hasegawa–Wakatani system}



\author{Luning Sun}
\email[]{sln91945@gmail.com}
\affiliation{Applied Materials, Inc., Santa Clara, CA, 95054, USA}
\author{Ben Zhu}
\email[]{ben.zhu@columbia.edu}
\affiliation{Department of Applied Physics and Applied Mathematics, Columbia University, New York, NY 10027, USA}
\affiliation{Columbia Fusion Research Center, Columbia University, New York, NY 10027, USA}
\author{Xin-Yang Liu}
\affiliation{University of Notre Dame, Notre Dame, IN 46556, USA}
\author{Deepak Akhare}
\affiliation{University of Notre Dame, Notre Dame, IN 46556, USA}
\author{Jian-Xun Wang}
\affiliation{Cornell University, Ithaca, NY 14853, USA}



\date{\today}

\begin{abstract}
Control of plasma turbulence remains a long-standing challenge in magnetically confined fusion research. Here, we explore a reinforcement-learning (RL) approach for bidirectional control of the turbulence-zonal-flow transition in the modified Hasegawa-Wakatani system, a minimal model of electrostatic drift-wave turbulence. Within our model, a weak zonal drag damps the otherwise long-lived zonal structures, giving them a finite lifetime and restoring the drive-damping balance required for repeatable transitions within finite control episodes; meanwhile, actuation is applied through a spatially distributed Gaussian source field under a time-weighted budget constraint on actuation costs. The plasma model is then coupled to CNN-based soft actor-critic and twin-delayed deterministic policy-gradient agents through a GPU-native JAX solver that is highly optimized for fast online training. In the turbulence-suppression task, the learned budget-aware schedule achieves the lowest time-integrated turbulent flux for a given consumed budget, outperforming both constant and linearly decreasing baselines across all tested unseen initial conditions. In the inverse zonal-break task, the agent discovers an up-down antisymmetric actuation pattern that induces radial $E\times B$ convection, disrupts the zonal structure, and sustains the turbulent state. A physics-informed warm-buffer initialization facilitates this discovery, as random exploration alone struggles to locate the narrow optimal manifold within the vast action space. These results demonstrate reinforcement learning as a practical trajectory optimizer for nonlinear plasma dynamics, such as turbulence control, and highlight the importance of physics guidance in such applications.
\end{abstract}

\pacs{}

\maketitle 


\section{Introduction}
\label{sec:intro}

Plasma turbulence remains a key mechanism for anomalous heat and particle transport in magnetically confined plasmas, dictating the overall confinement performance. Mitigating this transport in the core is essential for the economic viability of future burning plasma devices like SPARC and ITER. Conversely, heat exhaust presents a competing challenge where concentrated heat flux would damage plasma-facing components. Thus, in the boundary region, localized cross-field turbulence could actually be beneficial, as it broadens the heat-load footprint~\cite{brunner2018high}. Whether the objective is suppression in the core or enhancement at the edge, direct real-time control of turbulence is notoriously difficult due to the massive spatiotemporal scale separation between macroscopic machine parameters and microscopic turbulent eddies. Experimentally, turbulence control is often achieved indirectly. Macroscopic interventions, such as introducing external $E \times B$ flows via electrode biasing~\cite{schaffer1992effect, carter2009modifications} or specific RF heating schemes~\cite{ryter2014experimental}, can successfully enhance or suppress turbulence and trigger transitions. Typically, these experimental actuations rely on past experience or empirical scalings and simple, static control signals, such as applying a constant voltage to drive flow. Because magnetized plasmas are highly nonlinear and turbulence is a chaotic dynamical system, the performance of turbulence suppression or enhancement could be path-dependent. A static, constant drive may not be the most efficient approach; different dynamic control paths can lead to vastly different system states. Finding an optimal, non-trivial control trajectory through this chaotic landscape is an ideal challenge for reinforcement learning (RL), which excels at exploring high-dimensional actuator spaces to manage nonlinear dynamics.

Furthermore, machine-learning and artificial-intelligence techniques have already begun to accelerate fusion research: surrogate models predict disruptive instabilities~\cite{kates2019predicting}, reconstruct equilibria rapidly~\cite{sun2024impact}, restore degraded diagnostics via multimodal super-resolution~\cite{jalalvand2025multimodal}, extend stable operational spaces~\cite{kim2024highest}, and control detachment without in-situ diagnostics~\cite{zhu2025latent}. Among these, RL is arguably the most promising tool for \emph{direct} plasma control, with demonstrated magnetic shape and position control of core plasmas~\cite{degrave2022magnetic}, active tearing-instability avoidance~\cite{seo2024avoiding}, and trajectory design for disruption-free tokamak ramp-downs~\cite{wang2025learning}.
Applying RL directly to high-fidelity global turbulence models (5D gyrokinetic or 3D fluid) remains computationally prohibitive for online training. In fluid dynamics, however, RL has proven highly effective for controlling analogous nonlinear partial differential equations (PDEs). Deep RL has successfully learned closed-loop policies for spatiotemporal dynamics through online interaction, achieving wake stabilization, instability suppression, and flow regulation~\cite{bucci2019control, fan2020reinforcement, mondal2025shocks}; moreover, it has been extended to sensor placement and experimental flow control~\cite{watanabe2025effect}. Model-based and physics-informed formulations~\cite{liu2021physics} together with multi-fidelity and multi-agent strategies~\cite{bae2022scientific, sun2025multi} further improve scalability. Collectively, these results establish online RL as a viable pathway for controlling nonlinear PDE systems, and hence for reduced plasma-turbulence models such as the one studied here.

In this work, we explore RL for plasma turbulence control using a reduced-order model, the modified Hasegawa-Wakatani (MHW) system~\cite{numata2007bifurcation}, as a computationally tractable testbed.
While the original Hasegawa-Wakatani system and its variants~\cite{hasegawa1983plasma,biskamp1995nonlinear,numata2007bifurcation,dewhurst2009effects,majda2018flux} have been extensively studied to understand drift-wave turbulence and zonal-flow dynamics, control-oriented research using this model remains limited. Recently, these models have also served as targets for machine-learning-based surrogate modeling~\cite{gahr2024scientific,li2025reconstructing,vandewetering2026neural}, but direct control applications are still scarce. Classical reduced-order modeling techniques, such as Proper Orthogonal Decomposition (POD) and Dynamic Mode Decomposition (DMD), are well established for control in fluid and geophysical turbulence~\cite{rowley2005model}. In the plasma context, balanced truncation applied to the linearized HW system has achieved local stabilization with linear-quadratic controllers~\cite{goumiri2013reduced}. However, such classical approaches are inherently local and linear. They rely on stabilizing a limited number of unstable modes near a specific equilibrium and require the actuator to intervene sufficiently early. Consequently, these methods are not applicable to regulate fully developed, strongly nonlinear turbulent dynamics.
Furthermore, complete stabilization of turbulence and/or the underlying instabilities is often an unrealistic objective for practical fusion devices, since turbulent transport inevitably persists in most, if not all, confined plasmas. A more physically meaningful task is to control the turbulence by regulating its statistical properties, rather than attempting to completely suppress the driving instabilities. Achieving such optimal regulation is the goal of our study. This work is exploratory, but it pursues that goal \emph{directly}: the agent optimizes the actuation against the nonlinear turbulent transport itself, rather than against a proxy (e.g., a background gradient, a linear growth rate, or the stability of a linearized system) that need not reflect the saturated, nonlinear turbulent dynamics.

To achieve this, we first reformulate turbulence control within a revised MHW framework. By introducing a weak zonal-drag term to restore a generic drive-damping balance, we enforce the minimal structural requirement that zonal-flow-dominated regimes persist only under sustained drive~\cite{diamond2005zonal,chen2016physics}. Moreover, we consider a control source with a constrained \emph{budget} to reflect real experimental constraints, such as limited capacitor energy in biasing electrodes or finite-duration heating pulses in RF systems. This allows us to investigate optimal actuation policies under realistic physical constraints.
Within this modified model, we cast bidirectional turbulence control as a reinforcement-learning problem with a physics-based reward. First, in the suppression task, an agent learns to steer the system from a turbulent into a zonal-flow state more efficiently than trivial baselines at an equal consumed budget. Second, in the inverse task, an agent drives the system out of the zonal-flow state through an interpretable up-down antisymmetric pattern whose induced radial $E\times B$ convection breaks the zonal structure. Examining the latter task reveals that the effective actuation occupies a narrow, physics-selected manifold within the action space. Because of this restrictive reward-landscape geometry, random exploration struggles to find optimal policies, thereby motivating the use of a physics-informed warm-buffer initialization. This challenge highlights what we take to be the central difficulty of learning to control a turbulent system: optimizing a reward derived from a chaotic state, which suffers from substantial realization scatter and often features a nearly flat landscape away from the optimum. The remainder of the paper is organized as follows: Sec.~\ref{sec:problem} introduces the model and the control problem formulation, Sec.~\ref{sec:method} briefly describes the reinforcement-learning framework used in this study, Sec.~\ref{sec:result} presents the details of the results for the two control tasks and discusses the lessons learned during the study, and Sec.~\ref{sec:conclusion} summarizes our findings.

\section{Problem Formulation}
\label{sec:problem}

In this work, we adopt the MHW equations to capture the interaction between drift-wave turbulence and zonal flows, a key physical ingredient of realistic plasma-turbulence applications. The standard MHW formulation accumulates energy in the zonal modes and supports extremely long-lived, effectively absorbing zonal-flow-dominated states~\cite{numata2007bifurcation,grander2024hysteresis}. Once reached, these states can persist for times far exceeding any practical control horizon.
Since our goal is to study repeatable, bidirectional transitions within finite control episodes, we introduce a small linear drag on the zonal vorticity (i.e., the $-\gamma_\text{ZF}\bar{\varpi}$ term), which damps the zonal states and gives them a finite lifetime. Physically, this term mimics various collisionless and collisional dissipation mechanisms, such as magnetic pumping or ion-neutral friction, restoring a generic drive-damping balance to the system during the zonal-turbulence transition. In addition, we introduce an external actuation term to regulate the turbulence. The complete governing equations used in our study are thus given by
\begin{align}\label{eq:gov}
\frac{\partial}{\partial t}\varpi +[\phi,\varpi] &= \alpha(\tilde{\phi} - \tilde{n}) + \alpha\phi_\text{ext} - \gamma_\text{ZF} \bar{\varpi} -\mu\nabla_\perp^6\varpi\\
\frac{\partial}{\partial t}n + [\phi,n] &= \alpha(\tilde{\phi} - \tilde{n}) - \kappa\frac{\partial\phi}{\partial y} + \alpha\phi_\text{ext} - \kappa\frac{\partial\phi_\text{ext}}{\partial y} - \mu\nabla_\perp^6 n
\end{align}
where $n$, $\varpi=\nabla_\perp^2\phi$, and $\phi$ are the density, vorticity, and electrostatic potential, respectively. All quantities are expressed in the standard gyro-Bohm normalization (i.e., lengths in $\rho_s$, times in $\omega_{ci}^{-1}$); times quoted throughout are in these normalized units. The nonlinear $E\times B$ convection is represented by the Poisson bracket $[f,g] = \boldsymbol{\hat{z}}\cdot (\nabla f \times \nabla g)$, and the zonal and non-zonal components of a given quantity $f$ are defined as
\begin{equation}
    \bar{f}=\frac{1}{L_y}\int f dy,\quad \tilde{f} =f -\bar{f}.
\end{equation}
To ensure numerical stability at the grid scale, a sixth-order hyper-diffusion term is added to both equations.
The external actuation $\phi_\text{ext}$ enters as a prescribed control source whose structure follows previous control studies of this system~\cite{goumiri2013reduced}. Note that this represents a direct, defined actuation rather than a derived electrode response. We define $\phi_\text{ext}$ as a spatially distributed field composed of $N_A=4$ independent Gaussian actuators
\begin{equation}\label{eq:GaussField}
    \phi_\text{ext} = \sum_{i=1}^{4} a_i \cdot 2 \left(1 - \frac{r_i^2}{p^2}\right) \exp\left(-\frac{r_i^2}{p^2}\right)
\end{equation}
where $r_i^2 = (x - x_{\text{c},i})^2 + (y - y_{\text{c},i})^2$ and the spatial width is fixed at $p=5$. The actuator centers $(x_{\text{c},i}, y_{\text{c},i})$ are placed on a $2\times2$ grid at $L/4$ and $3L/4$ in both directions, allowing the RL agent to control a four-dimensional action vector $\mathbf{a}=(a_1,\dots,a_4)$ that sets the individual actuator amplitudes.

Although the layout uses only four actuators, allowing each amplitude to take either sign makes the forcing field highly expressive. Up to the symmetries of the square, the sign pattern of $(a_1,\dots,a_4)$ realizes five qualitatively distinct configurations: a uniform pattern (all four amplitudes sharing the same sign), a three-to-one split, and three two-to-two splits (adjacent row-wise, adjacent column-wise, and diagonal). The row-wise split, with opposite signs between the bottom and top actuator pairs, is the up-down antisymmetric pattern that the agent discovers in the zonal-break task (Sec.~\ref{sec:reward_hacking}).

\begin{figure}[!htb]
\centering
\includegraphics[width=\textwidth]{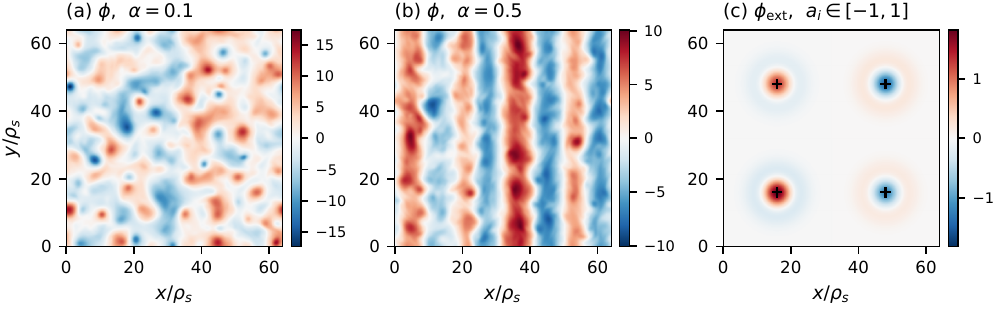}
\caption{Problem setup. Uncontrolled electrostatic potential $\phi$ in the two regimes studied, from our solver at $\kappa=1$ without actuation: (a) drift-wave turbulence at $\alpha=0.1$ and (b) a self-organized zonal flow at $\alpha=0.5$. (c) The actuator field $\phi_\text{ext}$: four Laplacian-of-Gaussian profiles [Eq.~\eqref{eq:GaussField}] centered on the $2\times2$ grid at $L/4$ and $3L/4$ (crosses), width $p=5$, shown for a random amplitude draw $a_i\in[-1,1]$.}
\label{fig:setup}
\end{figure}

To facilitate seamless integration with the reinforcement learning pipeline and to leverage hardware acceleration, we implemented the numerical solver for this model in Python using JAX. The spatial discretization employs the classic Arakawa scheme~\cite{arakawa1966computational} for the Poisson brackets to improve the long-term conservation of kinetic energy and enstrophy. Spectral methods are used for rapid field inversion (i.e., solving $\phi$ from $\varpi$) and for the exact evaluation of the hyper-diffusion terms, while a standard 4th-order central difference scheme is applied to the linear spatial derivatives. For time integration, the solver supports both a 2nd-order trapezoidal leapfrog scheme and a standard 4th-order Runge-Kutta (RK4) method. To ensure numerical fidelity, this JAX implementation was cross-benchmarked against the established Fortran code GDB~\cite{zhu2018gdb}, confirming the accuracy of the simulated physics (see Appendix~\ref{sec:appendix}).

Throughout our study, simulations are conducted on a $256\times256$ mesh over a $64\times 64$ doubly periodic domain, with the background density gradient fixed at $\kappa=1$. The hyper-diffusion coefficients are set to $\mu_\varpi=\mu_n=2\times10^{-5}$ in all evaluations reported here, providing the necessary grid-scale dissipation without artificially affecting the macroscopic turbulence and zonal-flow dynamics. Two distinct control tasks are investigated in this work, differentiated by the adiabaticity parameter. The first is a turbulence-suppression task at $\alpha=0.1$, which is close to the hydrodynamic limit where the uncontrolled system is dominated by strong drift-wave turbulence. The second is a zonal-break task at $\alpha=0.5$, where the uncontrolled system sustains a persistent zonal flow. Representative uncontrolled fields for the two regimes, together with the actuator layout, are shown in Figure~\ref{fig:setup}. In the turbulence-suppression task, the control policy is additionally subject to an actuation budget: a time-weighted cost on the actuation enters the reward, and a fixed cap $B_\text{total}$ on the total control effort is enforced by the environment. 
The complete per-task configuration is summarized in Table~\ref{tab:task_setup}.

\begin{table}[ht]
\centering
\caption{Per-task simulation and control configuration. Shared numerics:
$256\times256$ mesh, $64\times64$ doubly periodic domain, $\kappa=1$,
$\mu_\varpi=\mu_n=2\times10^{-5}$, actuator width $p=5$.}
\label{tab:task_setup}
\begin{tabular}{lcc}
\toprule
 & \textbf{Turbulence Suppression} & \textbf{Turbulence Enhancement} \\
\midrule
Adiabaticity $\alpha$            & $0.1$ & $0.5$ \\
Zonal drag $\gamma_\text{ZF}$    & $0.02$ & $5\times10^{-3}$ \\
Actuator layout                  & $2\times2$ ($N_A=4$) & $2\times2$ ($N_A=4$) \\
Action space                     & continuous, $a_i\in[0,8]$ & discrete, $a_i\in\{\pm3,\ldots,\pm9\}$ \\
Control interval $\Delta t_c$    & $0.1$ & $50$ \\
Episode length $T$               & $150$ & $100$ \\
Budget constraint                & Yes (Eq.~\ref{eq:reward_zonal}) & No \\
Reward                           & budget-aware flux (Eq.~\ref{eq:reward_zonal}) & scale-invariant flux \\
Warm buffer                      & $200$ episodes & $70$ episodes ($40$ antisym.\ $+$ $30$ random) \\
Algorithm                        & SAC & TD3 \\
\bottomrule
\end{tabular}
\end{table}

\begin{figure}[!htb]
\centering
\includegraphics[width=0.8\textwidth]{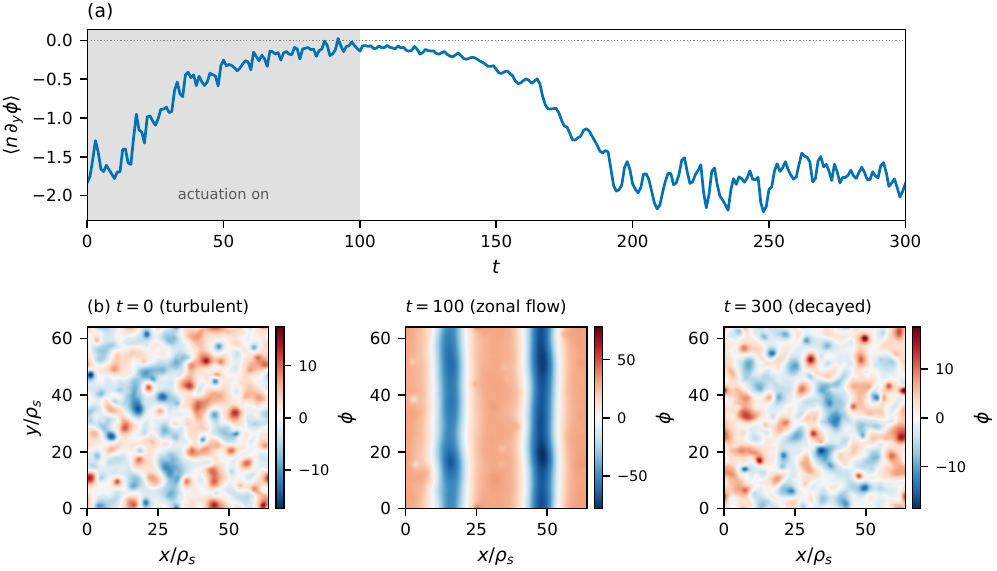}
\caption{Validation of the zonal drag for $\alpha=0.1$, $\gamma_\text{ZF}=2\times10^{-2}$: (a) domain-averaged flux metric $\langle n\,\partial_y\phi\rangle$ versus time; a constant actuation $a_i=8$ is applied over the shaded window and then turned off, after which the zonal state decays on the drag timescale; (b) potential $\phi$ at $t=0$ (turbulent), $t=100$ (zonal flow), and $t=300$ (back to turbulence).}
\label{fig:zonal_drag}
\end{figure}

The choice of $\gamma_\text{ZF}$ is verified prior to the control tasks to ensure that turbulence transitions can occur given the available actuation amplitude and within the control horizon. For example, at $\alpha=0.1$ (shown in Figure.~\ref{fig:zonal_drag}), starting from a fully developed turbulent state, a constant actuation $a_i=8$ applied over $t=0-100$ drives the domain-averaged flux metric $\langle n\,\partial\phi/\partial y\rangle$ (i.e., the transport measure used as the reward signal in Sec.~\ref{sec:method}) up toward zero, indicating the transition into a zonal-flow state. At $t=100$ the actuation is switched off, and under the zonal drag alone the zonal state decays back to turbulence on the order of the drag timescale, $\gamma_\text{ZF}^{-1}=50$. Together, these two phases confirm the intended drive-damping balance: zonal-flow states are reachable under a sustained drive and possess a finite lifetime without it. Hence, repeatable bidirectional transitions are accessible within a single control episode of length $T=150$.

\section{Methodology}
\label{sec:method}
\subsection{Overall Framework}
\begin{figure}[!htb]
\centering
\includegraphics[width=1.05\textwidth]{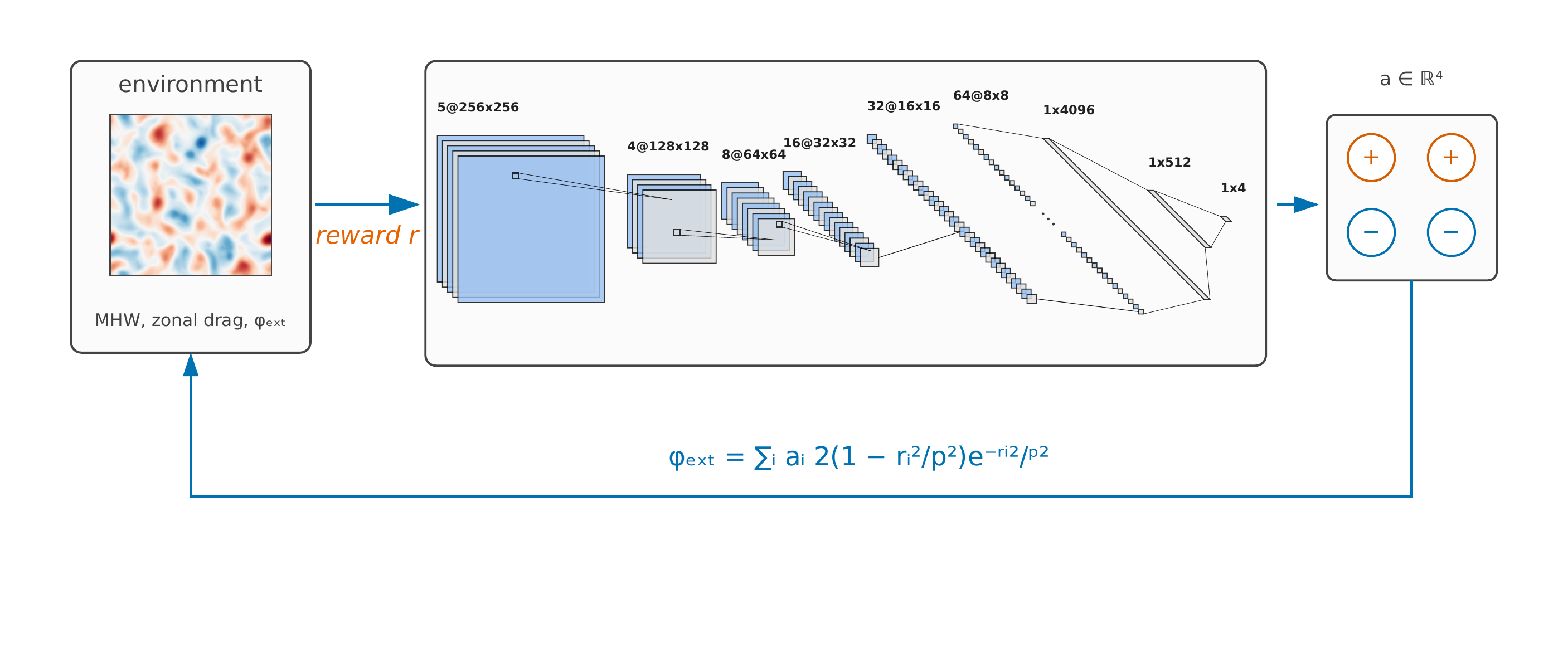}
\caption{Schematic of the control framework. The environment returns a five-channel $256\times256$ observation, comprising the potential $\phi$, the remaining budget, the zonal kinetic energy and its running integral, and the fraction of the episode remaining, together with the reward. A shared convolutional encoder feeds an actor head and twin critic heads, the latter fusing the action and remaining budget. The four actuator amplitudes set $\phi_\text{ext}$, closing the loop.}
\label{fig:framework}
\end{figure}

We adopt a model-free, online RL framework that learns an optimal control policy $\pi$ by interacting directly with the environment through trial-and-error search (Figure.~\ref{fig:framework}). The environment is the JAX-based solver, jaxHW, developed for this work. We chose to build this solver because established legacy codes, such as GDB~\cite{zhu2018gdb}, are built for physics fidelity rather than the many-query rollouts required for online training. On a Perlmutter A100 GPU, jaxHW advances a $256\times256$ MHW simulation at $0.36$~ms per step, which is roughly an order of magnitude faster than GDB ($4.16$~ms per step on CPUs). Equally importantly, it runs end-to-end on the GPU alongside the agent networks. Solver verification and per-step timings are detailed in Appendix~\ref{sec:appendix}. The control agent uses convolutional neural network (CNN) layers to extract feature maps from a five-channel, $256\times256$ observation space. This space consists of the observed potential $\phi$, the remaining budget, the zonal kinetic energy, its running integral, and the fraction of the episode remaining. These extracted features are then mapped to the four-dimensional action vector. The density and vorticity are advanced by the solver and enter the reward, but are not passed to the agent. As for RL framework, we employ two actor-critic algorithms, Soft Actor-Critic (SAC)~\cite{haarnoja2018soft} and Twin-Delayed Deep Deterministic Policy Gradient (TD3)~\cite{fujimoto2018addressing}. 

\subsection{Physics-based Reward Function}
In the first task, the control goal is to transition to a zonal-flow state as quickly as possible and to maintain the zonal-flow statistics, all while adhering to a fixed budget constraint. Therefore, we define the physics-based reward function as:

\begin{equation}
r = \min\!\left(\left\langle n \,\frac{\partial\phi}{\partial y}\right\rangle, 0\right) - \lambda_1 \cdot \frac{\sum_i |a^{RL}_i|}{4 \times a_{norm}} \cdot 50^{t/T} - \lambda_2 \left\| a^{RL} - a^{exec} \right\|^2
\end{equation} \label{eq:reward_zonal}
with
\begin{align}
a^{{exec}} = \begin{cases}
a^{{RL}}, & \text{if } \sum_i |a^{{RL}}_{i}| \cdot \Delta t \leq B_{\text{rem}} \\
a^{{RL}} \cdot \dfrac{B_{\text{rem}}}{\sum_i |a^{{RL}}_{i}| \cdot \Delta t}, & \text{otherwise}
\end{cases}
\end{align}
where $t$ is the time within the control episode of length $T=150$ (i.e., $N_c=T/\Delta t_c=1500$ control steps), $\lambda_1 = 0.01$, $\lambda_2=0.05$, and $a_{norm}=8$ is the actuator amplitude bound. The total budget is $B_\text{total}=\sum_i |a^{RL}_i|\Delta t_c$, $B_{\text{rem}}$ is the remaining budget, and $\Delta t = \Delta t_c = 0.1$.
The three terms in the reward function are, respectively, the (clipped to be non-positive) flux reward, a time-weighted action-cost penalty that increasingly discourages late-episode spending, and an overspend penalty that flags actions exceeding the remaining budget as infeasible.

The flux term deserves comment, as it anchors the reward in transport physics. In the HW normalization the radial $E\times B$ particle flux is $\Gamma_n=-\kappa\langle n\,\partial\phi/\partial y\rangle$ (this sign convention is used consistently throughout). Thus, the domain averaged metric $\langle n\,\partial\phi/\partial y\rangle<0$ corresponds to down-gradient transport driven by active turbulence, while a saturated zonal-flow state suppresses the flux toward zero. The clip $\min(\cdot,0)$ discards transient up-gradient excursions, which are not physically meaningful targets. Because the RL return sums the reward over the episode, the agent effectively minimizes the \emph{time-integrated} turbulent transport. This represents the cumulative quantity a confinement device actually pays for, rather than the flux at any single instant. This choice also provides a dense per-step learning signal, which eases credit assignment compared with a sparse terminal reward on the final state. It is thus more robust to the burstiness (or, intermittence) of the instantaneous flux: rewarding instantaneous values alone would credit transient dips that do not reflect a genuine approach to the zonal state. 

For the zonal-break task, where the objective reverses, the same flux is used but it is normalized by the zonally averaged potential amplitude into a scale-invariant form. The reward-landscape analysis motivating this formulation is presented in Sec.~\ref{sec:reward_hacking}.

\subsection{Training and Replay-Buffer Setup}

The control policy $\pi(\boldsymbol{u}^o;\boldsymbol{\theta}_\pi)$, parameterized by network weights $\boldsymbol{\theta}_\pi$, maps an observation $\boldsymbol{u}^o=\mathcal{O}(\boldsymbol{u})$ of the system state $\boldsymbol{u}$ to the action vector $\boldsymbol{a}$. In this work $\mathcal{O}$ is the identity, so the agent observes the full $256\times 256$ fields. Training maximizes the expected discounted return $R=\sum_i \gamma^{i-1} r_i$ (discount factor $\gamma$) accumulated over a control episode. We use two standard off-policy actor-critic algorithms: SAC~\cite{haarnoja2018soft} and TD3~\cite{fujimoto2018addressing}.

A practical advantage of off-policy algorithms is that the replay buffer can be generated once, offline, and reused across hyperparameter searches, avoiding a fresh set of environment rollouts for every configuration. We exploit this by pre-computing a warm buffer offline: $200$ episodes under uniform-random actions for the turbulence-suppression task, and $70$ episodes for the zonal-break task. For the latter, $40$ episodes use antisymmetric actions and $30$ are fully random (Sec.~\ref{sec:reward_hacking}). Because the observations inculde full $256\times256$ $\phi$ field stacks, the buffer is stored as memory-mapped arrays alongside the transition record. This ensures that repeated hyperparameter runs stream the same cached rollouts from disk instead of regenerating them.

\section{Results}
\label{sec:result}
With the problem formulated and the RL framework established, we now investigate two complementary control tasks that address opposite directions of the turbulence-zonal-flow transition. The first task is turbulence \emph{suppression}, which drives a turbulence-dominated plasma into a quiescent zonal-flow state, an analogue of reducing anomalous transport for confinement. The second is the inverse turbulence \emph{enhancement}, or zonal-break, task: disrupting a self-sustained zonal flow to restore cross-field transport, an effect that is desirable near the plasma edge to broaden the heat-load footprint. Both tasks utilize the same actuator formulation.

\subsection{Turbulence Suppression}
\label{sec:control}
In the hydrodynamic regime (i.e., $\alpha=0.1$, $\gamma_\text{ZF}=0.02$), the uncontrolled system is dominated by drift-wave turbulence. The agent's objective is thus to suppress the turbulent transport by driving the plasma into a zonal-flow state within a fixed actuation budget. To achieve this, we train a SAC agent~\cite{haarnoja2018soft} at a control interval $\Delta t_c=0.1$ over a horizon $T=150$ ($1500$ control steps), with actuator amplitudes bounded to $a_i\in[0,8]$. 

Because the optimization target is minimizing the accumulated turbulent flux, we want to explore whether a learned control policy can outperform two heuristic baselines: a constant drive at $a_i=4$ and a linear ramp-down from $a_i=8$ to $0$ for all 4 actuators. Both baselines operate under the same budget constraint, $B_\text{total}=2400$.

As discussed in the previous section, our reward is $\left\langle n \,\frac{\partial\phi}{\partial y}\right\rangle$ and the turbulent particle flux in this system is $\Gamma_n=-\kappa\langle n\,\partial\phi/\partial y\rangle$. Therefore, a reward approaching zero corresponds to the turbulent transport being driven down toward its quiescent zonal-flow level, and the time integral of the reward (i.e., the total return) measures the cumulative transport incurred over the control window.

\begin{figure}[!htb]
\centering
\includegraphics[width=0.8\textwidth]{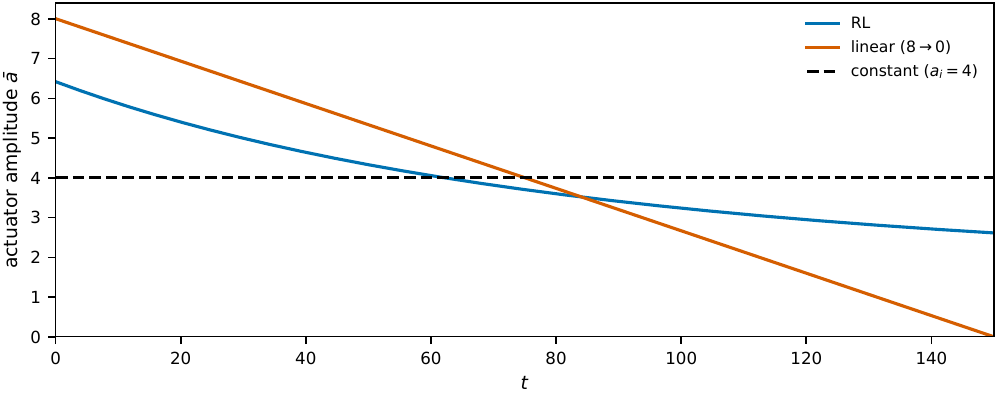}
\caption{Actuation schedules for the suppression task: amplitude averaged over the four actuators versus control step, for the learned policy (blue), the constant baseline ($a_i=4$), and the linear-decrease baseline ($8\to0$). The learned schedule is identical across the five initial conditions and across the four actuators.}\label{fig:act}
\end{figure}

\begin{figure}[!htb]
\centering
\includegraphics[width=0.8\textwidth]{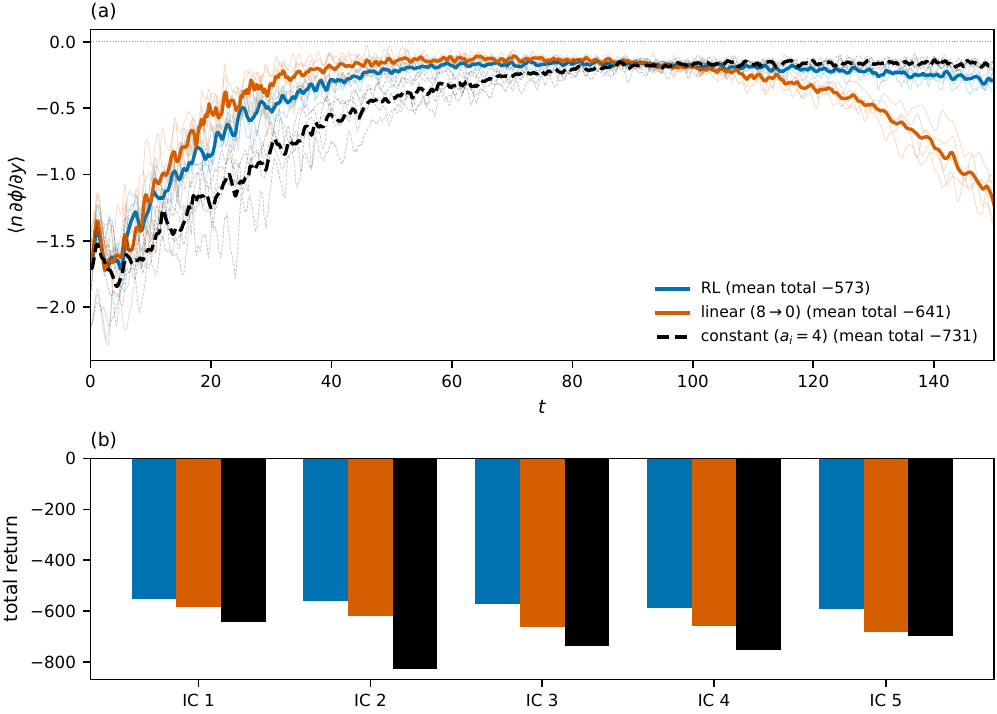}
\caption{Turbulence-suppression task: (a) per-step flux reward $\langle n\,\partial\phi/\partial y\rangle$ versus control step for the learned RL policy (blue), the linear-decrease baseline (red), and the constant baseline (black dashed); thin lines are the five unseen initial conditions, thick lines their means, and the legend quotes each method's mean total return; (b) total episode return per initial condition.}\label{res:contrl_res}
\end{figure}

The two heuristic control schedules are shown in Figure~\ref{fig:act}, and their effects on five random initial turbulent states are shown in Figure~\ref{res:contrl_res}. The linear ramp-down hits the system hard and quickly reduces the flux, yielding a larger instantaneous reward. However, it fails to hold this state because the actuation strength becomes too weak at later times, yielding a smaller instantaneous reward. In contrast, the constant policy gradually reduces the flux, taking longer to reach its maximum reward, but maintains this state until the end of the control period. Intuitively, one would expect a \emph{better} policy to exist that combines the strengths of these two heuristic schedules, namely starting with strong actuation and then reserving enough budget to maintain a finite actuation level until the end. However, the exact form of such a schedule is non-trivial, as the chaotic nature of the system dictates a highly nonlinear response to the actuation.

Guided by this intuition, and exploiting the off-policy structure of our algorithm, a warm buffer of $150$ episodes/$30{,}000$ steps is pre-computed offline and reused across parameter searches, followed by $50{,}000$ training steps. The agent acts through a prescribed base schedule rather than in isolation: its output is a bounded deviation, $a(t)=a_\text{base}(t)+\sigma(\mu)\,\Delta_\text{max}$. Here $\Delta_\text{max}=0.8$, $\sigma$ is a softsign squashing function, and the deviation is applied to a linear reference ramp $a_\text{base}: 7.0\to2.4$. 

The learned policy, shown in Figure~\ref{fig:act}, matches our expectations. It is \emph{front-loaded}: it applies strong forcing early to trigger the turbulence-to-zonal-flow transition, then tapers along a non-trivial, nonlinear profile as the emerging zonal flow becomes self-sustaining against the weak zonal drag. This creates a schedule that neither heuristic baseline reproduces and that is not obvious a priori. 
Note that this task exercises both temporal and spatial degrees of freedom: the agent sets the amplitude of each of the four actuators at every control step freely. However, in the spatial dimension, the four amplitudes converge to nearly the same value, making the non-trivial content of the policy purely its temporal trajectory. This is somewhat expected as the four actuators are equivalent under the discrete symmetry of the $2\times2$ array on the doubly periodic domain, and the turbulent ensemble is statistically homogeneous, so the budget-optimal forcing is distributed symmetrically.

The learned RL policy was then evaluated against the two heuristic baselines (all three consume the same total actuation to within $1\%$, $B_\text{total}\approx2400$) on unseen turbulent initial conditions drawn from a held-out pool. 
Across the five representative initial conditions shown in Figure.~\ref{res:contrl_res}, the RL policy attains the highest (least negative) total return of the three methods (mean $-573$, versus $-641$ for the linear-decrease and $-731$ for the constant baseline) and leads on each of the five. 
As anticipated, the two baselines fail in complementary ways against the RL policy: the linearly decreasing baseline drive is competitive in total return but relaxes too early and loses control of the flux late in the episode, whereas the constant baseline drive holds the late-time state yet spends its budget inefficiently and yields the lowest total return. The learned policy outperforms both by allocating actuation \emph{when} it is most effective rather than at a fixed rate.

\begin{figure}[!htb]
\centering
\includegraphics[width=0.8\textwidth]{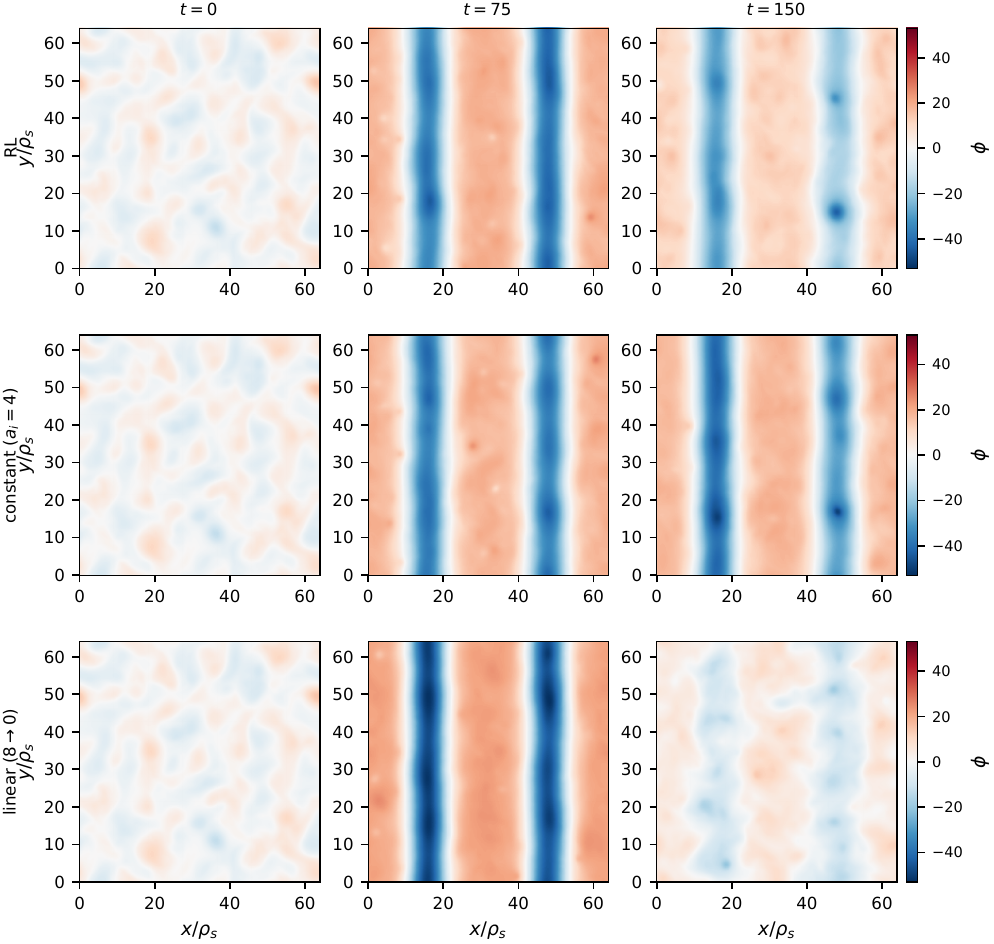}
\caption{Electrostatic potential $\phi$ at $t=0$, $75$ and $150$ (columns) under the RL policy, the constant drive and the linearly decreasing drive (rows), for a representative unseen initial condition. The RL policy organizes the field into coherent zonal bands and retains them to the end of the episode, whereas the linearly decreasing drive loses the zonal state late.}\label{fig:field_snap_suppress}
\end{figure}

The typical corresponding field evolution is shown in Figure.~\ref{fig:field_snap_suppress}: under the RL policy the turbulent field organizes fastest into coherent zonal bands and retains them through the end of the episode. Furthermore, in this particular setup, the RL schedule performs very similarly across the five independent initial conditions, evidencing a robust and physically interpretable strategy. This robustness is by design.
A policy trained on a single initial condition overfits to it and transfers poorly to others; however, training across an ensemble, together with a CNN encoder that compresses the state into a low-dimensional representation, discards initial-condition-specific details. This removes the policy's initial-condition sensitivity, yielding a policy that generalizes across unseen states. The same representation, however, also collapses the state dependence that a genuine feedback control loop would exploit. The tight consistency is thus double-edged, indicating that the policy has converged to a near-open-loop schedule rather than a state-feedback law, a point we should like to make explicit.

\begin{figure}[!htb]
\centering
\includegraphics[width=0.7\textwidth]{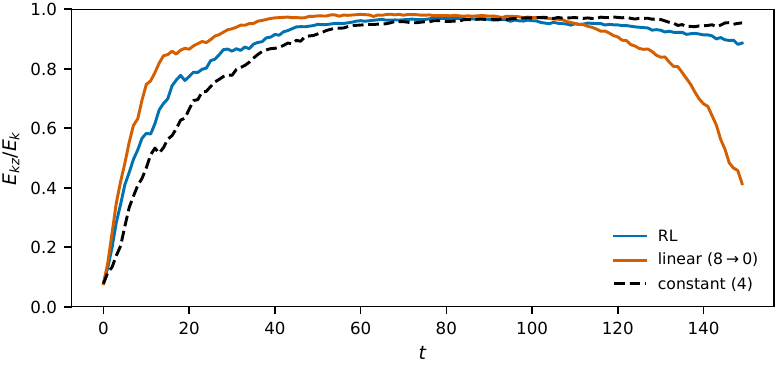}
\caption{Zonal-flow energy fraction $E_{kz}/E_k$ versus time for the learned policy and the two baselines, computed from the recorded $\phi$ fields of the controlled trajectories. The linear ramp attains the highest peak but relinquishes the zonal state late in the episode.}
\label{fig:ezf}
\end{figure}

To establish that the controlled system genuinely reaches a zonal-flow state, rather than merely hacking the reward signal, we examine a reward-independent diagnostic: the zonal-flow energy fraction $E_{kz}/E_k$, where $E_{kz}=\tfrac{1}{2}\langle(\partial_x\bar\phi)^2\rangle$ is the kinetic energy of the zonally averaged flow and $E_k=\tfrac{1}{2}\langle|\nabla\phi|^2\rangle$ the total kinetic energy. Both follow from the recorded $\phi$ fields, so the diagnostic is evaluated on the same trajectories that produced Figure.~\ref{res:contrl_res}. Figure~\ref{fig:ezf} shows the results. All three policies drive the fraction from its turbulent level ($0.08$) toward unity, confirming that the transition is physical and not an artifact of the reward. The policies are clearly distinguished by what happens late in the episode: the linear ramp reaches the highest peak ($0.98$) but then loses the state as its authority decays, ending at $0.41$, whereas the learned policy holds $0.89$ and the constant drive maintains $0.96$. This reflects the same complementary failure seen in the flux traces, now visible in a purely structural measure. It clearly demonstrates why the learned schedule wins on time-integrated transport: it reaches the zonal state early, like the linear ramp, and retains it, like the constant drive.

\subsection{Turbulence Enhancement: the Zonal-Break Task}
\label{sec:reward_hacking}
Encouraged by the success of the turbulence-suppression task, we now apply RL to the opposite task. In the more adiabatic regime (e.g., $\alpha=0.5$, $\gamma_\mathrm{ZF}=5\times10^{-3}$), the uncontrolled system settles into a persistent zonal flow, and the control objective is reversed: starting from a stable zonal-flow state, the agent must learn an actuation policy that disrupts the zonal structure and enhances cross-field turbulent transport.

In this $2\times 2$ actuation setup, it is quite straightforward to envision what the optimal actuation should look like. Because the radial particle flux is carried by the radial $E\times B$ drift $\tilde{v}_x=-\partial_y\tilde{\phi}$, only potential structure that varies \emph{along} the flux-surface direction ($y$) produces radial convection. Hence, an actuation pattern with opposite signs between the top and bottom actuator pairs, i.e., an up-down antisymmetric configuration, imposes a potential difference along $y$. This creates an electric field along the flux surface, whose $E\times B$ response is a coherent radial flow that carries density across the zonal bands and, through the $-\kappa\,\partial_y\phi_\text{ext}$ term in Eq.~\eqref{eq:gov}, taps the background gradient directly as a flux drive. A left-right antisymmetric pattern, by contrast, imposes a \emph{radial} electric field, whose $E\times B$ response is a sheared \emph{poloidal} flow which further drives the system into a deep-zonal-flow state. A uniform (same sign) setup also pushes the system to turbulence suppression (demonstrated in the previous task).

This task is deliberately a search over near-static actuation \emph{spatial} configurations rather than temporal control. With only $N=2$ control steps at a long control period $\Delta t_c = 50$, what is optimized is the spatial sign pattern and amplitude of the four actuators. We prepare a pool of five natural zonal-flow initial conditions by evolving the uncontrolled system until the turbulent flux $\langle n\,\partial\phi/\partial y\rangle$ saturates. The four-actuator $2\times2$ layout is retained with discrete amplitude levels $|a_i| \in [ 3\,\ldots,\,9]$ (small values excluded to prevent trivial inaction). Given the limited number of actions available, we originally expected the RL agent to easily rediscover this optimum. However, it proved to be much more challenging than the turbulence-suppression task.  

\begin{figure}[!htb]
\centering
  \includegraphics[width=0.8\textwidth]{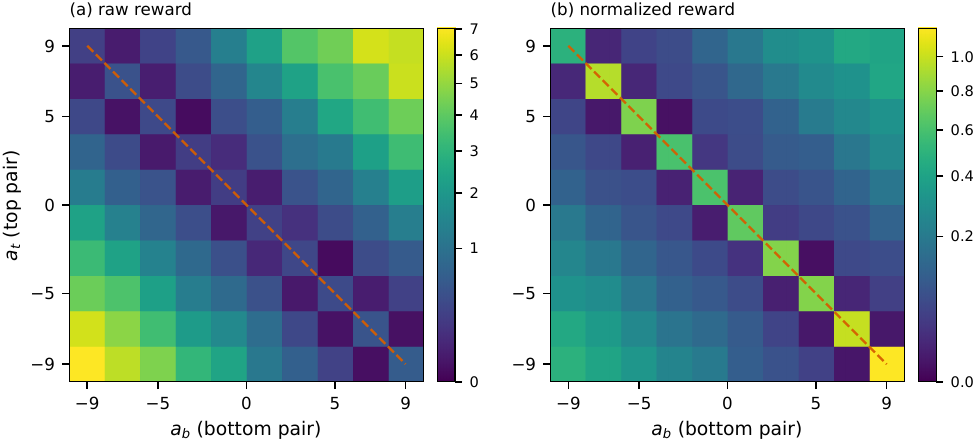}
  \caption{Action-space sweep with fixed pairwise actions where $a_b$ and $a_t$ are the bottom- and top-pair amplitudes, over a $10\times10$ grid; the dashed line marks the antisymmetric locus $a_t=-a_b$. (a)~Under the raw reward, same-sign corner-actions dominate, masking the anti-diagonal and indicating susceptibility to reward hacking. (b)~Under the normalized reward, Eq.~\eqref{eq:reward_norm}, the antisymmetric anti-diagonal is the true high-reward region.}
\label{fig:breakzonal_4_combined}
\end{figure}

Using the same reward function as in the suppression task, the RL agent always converges on a uniform setup, i.e., either $a_i\approx 9$ or $a_i\approx -9$ for all four actuators, which we know is a physically incorrect setup. This is because the raw reward, or \emph{turbulent} flux can be inflated by driving $\phi$ to extremely large amplitudes without genuine turbulent fluctuation. 
Figure~\ref{fig:breakzonal_4_combined}(a) shows an action-space sweep under the raw reward $r = \max(-\langle n\,\partial\phi/\partial y\rangle_\tau,\,0)$. The highest scores appear in the corners where $a_t$ (the top-pair amplitude) and $a_b$ (the bottom-pair amplitude) share the same sign and large amplitude. In these cases, the external forcing inflates $\phi$ far beyond the natural turbulence range, producing artificially large flux through the $\partial\phi/\partial y$ term without genuinely disrupting the zonal structure. An RL agent optimizing this raw reward thus exploits this shortcut rather than learning physically meaningful control; this is a classic scenario of \emph{reward hacking}.

To resolve this, we therefore adopt a scale-invariant normalized reward
\begin{equation}\label{eq:reward_norm}
r = 10\max\!\left(-\left\langle n\,\frac{\partial\phi}{\partial y}\right\rangle_\tau \Big/ C_\phi,\;0\right),
\end{equation}
where $\langle\cdot\rangle_\tau$ denotes a causal 20-step moving average of the instantaneous flux and $C_\phi = \langle\,|\bar\phi|\,\rangle_x$ is the spatially-averaged amplitude of the zonally-averaged potential. This normalization largely cancels the artificial gain from $\phi$ inflation as shown in Figure~\ref{fig:breakzonal_4_combined}(b). The anti-diagonal ridge $a_t = -a_b$ emerges as the sole high-reward region, with the peak at $(a_b, a_t) = (+9,-9)$ reaching a mean return of $1.18$, approximately $20\times$ higher than typical off-diagonal cells. Note that variation along this ridge is largely driven by initial-condition scatter; the two extreme polarities are physically equivalent up to background-gradient asymmetries.
Furthermore, we train a TD3 agent using the same CNN architecture as in Section~\ref{sec:control}. TD3 replaces the SAC agent used in the suppression task because the optimum of this task lies at the boundary of the action space, namely, the maximum-amplitude antisymmetric pattern. SAC's entropy-regularized stochastic policy is biased toward the interior and reaches such boundary optima only slowly, whereas the deterministic TD3 policy attains them directly.

Despite this improvement, the agent still struggled to find the optimal configuration. We find that this is due to the extreme narrowness of the optimal configuration within the action space, combined with the inherently noisy nature of the turbulent system (where gradients near the optimum can be largely obscured by turbulent fluctuations). While Figure~\ref{fig:breakzonal_4_combined} reduces the problem to 2D for better illustration, in the full four-dimensional action space, this effective antisymmetric manifold is vanishingly small, making it virtually impossible to discover by random exploration alone. Therefore, we apply a warm buffer of 70 offline-precomputed transitions, with $\geq 50\%$ of the cases set to antisymmetric configurations to encourage the agent to explore near this neighborhood. Specifically, we use 40 episodes with up-down antisymmetric actions $[-a,\,-a,\,+a,\,+a]$ at amplitudes $a\in\{3,\ldots,9\}$ drawn uniformly, plus 30 episodes with fully random independent actuator values. With this physics-informed guidance, the agent ultimately rediscovered the optimal configuration successfully.

\begin{figure}[!htb]
\centering
\includegraphics[width=0.8\textwidth]{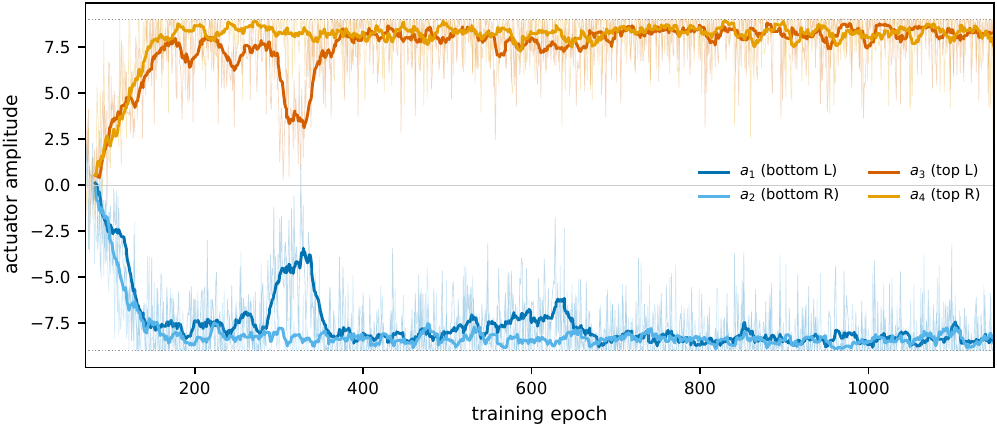}
\caption{Convergence of the four actuator amplitudes during training of the zonal-break agent; light traces are per-epoch network outputs, heavy traces their running mean. The antisymmetric sign pattern, bottom pair negative and top pair positive, is acquired within the first few hundred epochs, after which the amplitudes drift toward the boundary of the action set (dotted lines at $\pm9$).}\label{fig:breakzonal_2}
\end{figure}

Figure~\ref{fig:breakzonal_2} traces how the policy arrives there. With the warm buffer, the antisymmetric \emph{sign} pattern is acquired early, holding on $88\%$ of epochs below epoch $200$ and on every epoch thereafter, while the \emph{amplitude} drifts upward over training toward the boundary of the action set, from a mean magnitude of $5.6$ over the first $200$ epochs to $8.3$ over the last $200$. The deployed checkpoint applies the same constant $[-9,\,-9,\,+9,\,+9]$ configuration on every initial condition, that is, open loop in practice. The early acquisition of the sign pattern reflects the antisymmetric warm buffer, which seeds the search near the physically favored family; what training adds is the amplitude. The pattern applies negative forcing on the bottom actuator pair and positive forcing on the top pair, imposing exactly the along-flux-surface potential difference anticipated above: the resulting radial $E\times B$ convection carries density across the zonal bands and re-establishes the radial flux. 

\begin{figure}[!htb]
\centering
\includegraphics[width=0.8\textwidth]{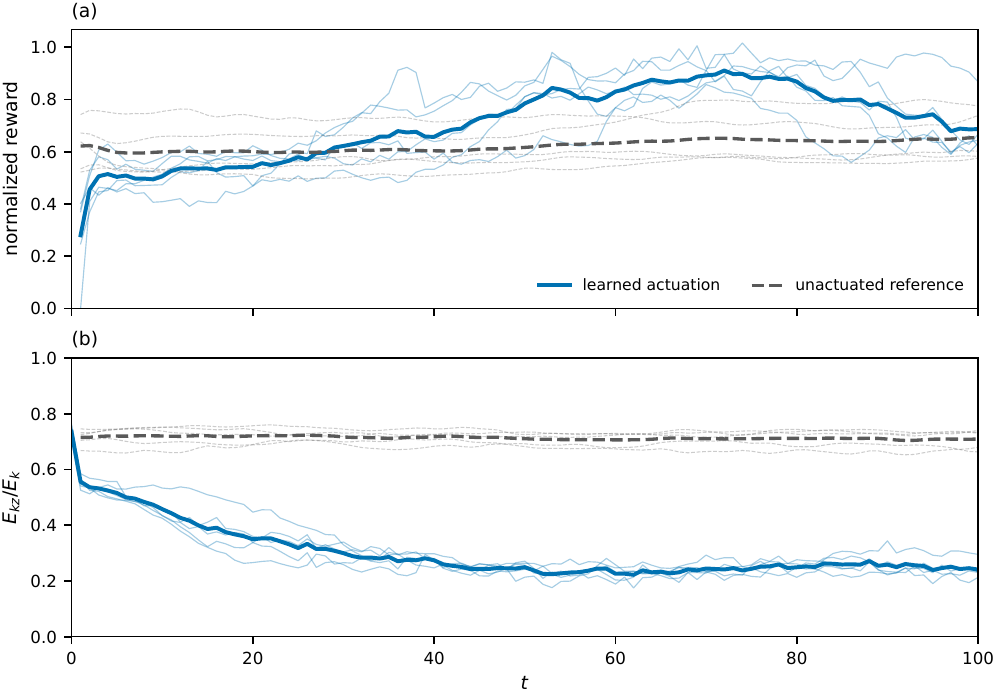}
\caption{Zonal-break task on the five evaluation initial conditions. (a) Normalized reward and (b) zonal-flow energy fraction $E_{kz}/E_k$ versus time under the learned actuation (blue; thin lines individual initial conditions, thick line their mean), compared with the unactuated evolution of comparable natural zonal-flow states (grey dashed,  at the same parameters). The reward is scale-invariant and returns toward the unactuated level late in the episode, while the energy fraction separates the two cases throughout.}\label{fig:breakzonal_1}
\end{figure}

Figure~\ref{fig:breakzonal_1}(a) shows the normalized reward, evaluated with $\Delta t_\mathrm{eval}=1$ sub-steps while keeping the policy cadence at $\Delta t_c=50$. Under the learned actuation the reward climbs from $\approx0.1$ at $t=0$ to a peak near $0.9$ at $t\approx70$--$80$, giving a mean total return of $70.6$ across the five initial conditions (range $63$ to $77$), and remains elevated through the end of the episode.
The normalized reward alone, however, may understate the effect ($0.62$ uncontrolled v.s. $0.67$ controlled, a difference of only $7\%$), because it is scale-invariant by construction: dividing by $C_\phi$ removes precisely the amplitude growth that the actuation produces, so the late-episode reward relaxes back toward the level of an unactuated zonal state. The structural diagnostic separates the two cases unambiguously [Fig.~\ref{fig:breakzonal_1}(b)]. Under actuation $E_{kz}/E_k$ falls from $0.74$ to $0.24$, and the zonal share of the $\phi$ variance from $0.96$ to $0.61$ as the field amplitude grows several-fold, whereas the unactuated states hold $E_{kz}/E_k\approx0.71$ throughout. The actual transport response follows the same separation. In unnormalized terms, unactuated evolution of comparable natural zonal states keeps the domain-averaged flux at $\langle n\,\partial\phi/\partial y\rangle=-0.22$, against $-0.40$ under the learned actuation, an enhancement by a factor of $1.8$. 
Together, these measures place the end state between an intact zonal flow and fully developed turbulence: the coherent bands are broken and cross-field transport is enhanced, but a large-scale zonal component survives at the end of the episode. Since the reward for this task is the normalized flux, these energy measures serve strictly as independent diagnostics.

\begin{figure}[!htb]
\centering
\includegraphics[width=0.8\textwidth]{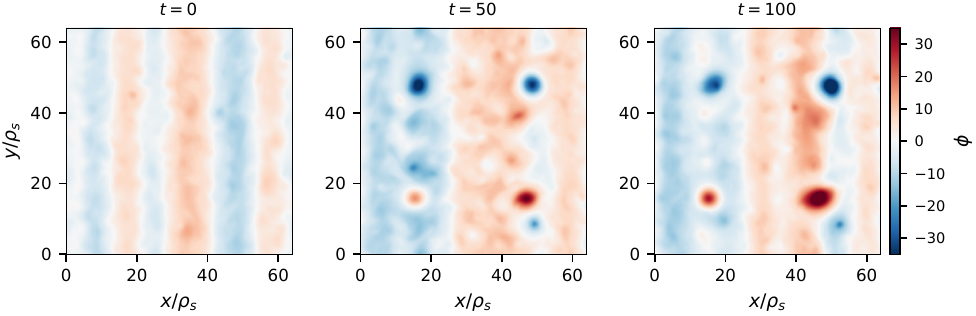}
\caption{Snapshot of $\phi$ at $t=0$, $50$ and $100$ for IC~0 under the learned configuration, showing the break-up of the coherent zonal bands. The zonal share of the $\phi$ variance falls from $0.96$ to $0.61$ while the field amplitude grows several-fold, so the end state is a strongly perturbed zonal flow rather than developed turbulence. The actuator sites appear as localized vortices of sign opposite to the applied amplitude.}\label{fig:breakzonal_3}
\end{figure}

Figure~\ref{fig:breakzonal_3} shows the $\phi$ field at $t=0$, $50$, and $100$ for a representative initial condition. At $t=0$ the field displays the ordered zonal bands characteristic of a zonal-flow state. By $t=50$ and $t=100$, the forcing has introduced irregular streaks and partially broken the large-scale coherence. To avoid confusion when interpreting these fields near the actuators, one detail is worth stating explicitly, because it governs how the actuation is read off the visualizations: the potential response $\phi$ carries the \emph{opposite} sign to the applied amplitude $\phi_\text{ext}$. Because the actuation enters the vorticity equation, recovering $\phi$ from $\varpi=\nabla_\perp^2\phi$ reverses the sign of a localized structure. Consequently, the positively forced pair appears as negative $\phi$ lobes, and the negatively forced pair appears as positive ones.

\subsection{Discussion}
\label{sec:discussion}

Several points deserve discussion beyond the individual tasks. First, the zonal-drag coefficient is not merely a numerical convenience but the knob that sets the difficulty of the control problem. The drag damps the zonal state, giving it a lifetime of order $\gamma_\text{ZF}^{-1}$. For example, in the validation test shown in Fig.~\ref{fig:zonal_drag}, this lifetime is roughly $80$ time units, which is comparable to, yet shorter than, the episode length $T=150$. This finite lifetime ensures that the control task is neither trivial (where an immortal zonal state would require no further actuation once reached) nor hopeless (where a drag much stronger than the drive would erase the target state faster than the budget allows). The per-task values in Table~\ref{tab:task_setup} sit in this intermediate regime.

Second, the learned policies operate in the strongly nonlinear response regime. The actuator amplitudes $a_i\lesssim 8$ produce an external potential whose peak magnitude is comparable to the saturated zonal potential itself, and $\phi_\text{ext}$ enters the dynamics through the same channels as the physical potential. This is precisely the regime that linear control approaches~\cite{goumiri2013reduced} cannot address. In experimental terms, $\phi_\text{ext}$ is the idealized analogue of a localized potential source, such as a biasing electrode driving $E\times B$ shear~\cite{schaffer1992effect,carter2009modifications}. Correspondingly, the hard actuation budget mirrors the finite capacitor energy or pulse duration of such systems.

Third, the two tasks jointly illustrate a lesson we expect to be generic for turbulence control: while the physics-based reward function is intuitive and straightforward, it may not work well off the shelf.  For instance, in the zonal-break task, the optimum is the up-down antisymmetric pattern anticipated by the flux-geometry argument. However, this optimal pattern only occupies a narrow manifold of the action space. Furthermore, the action-space sweep (Fig.~\ref{fig:breakzonal_4_combined}) shows that the reward landscape near that manifold is essentially flat or, even worse, deviates to a wrong ``optimum". This restrictive geometry defeats random exploration and makes a physics-informed seeding of the replay buffer operationally necessary rather than merely helpful. Because the $E\times B$ geometry that selects this pattern is common to all strongly magnetized plasmas, we expect warm-starting from physics-classified actuation patterns to transfer to more realistic settings.

Finally, we must note that this study is exploratory, which delimits the scope of the evidence assembled here. Each task was evaluated at a single budget level, a single zonal-drag value, and a single training run per algorithm, with robustness probed across ensembles of unseen initial conditions rather than through hyperparameter scans. Within this scope, however, the conclusions are consistent and physically interpretable.

\section{Conclusion}
\label{sec:conclusion}

We present a reinforcement-learning study for the bidirectional control of the transition between drift-wave turbulence and zonal flows in a modified Hasegawa--Wakatani system. Two physically motivated elements make the control problem well posed. Frist, a weak zonal-drag term $-\gamma_\text{ZF}\bar{\varpi}$ damps the otherwise long-lived, effectively absorbing zonal states of the MHW system. This gives them a finite lifetime, ensuring that turbulence-zonal transitions remain accessible within finite control episodes. Second, a spatially distributed Gaussian actuator field under a time-weighted actuation cost reflects the finite control budget of realistic actuation schemes. The many-query training that these tasks demand is made practical by a GPU-native JAX solver, jaxHW,  cross-benchmarked against the GDB code.

Two control tasks were solved using SAC/TD3 agents and a physics-based, budget-aware reward. In the turbulence-suppression task, the learned policy identifies a decreasing actuation schedule. This schedule is distinct from the heuristic ramp-down ($8\to 0$) and attains the lowest time-integrated transport among the three methods at equal consumed budget. It outperforms the baselines on all five unseen initial conditions tested. The advantage is not the speed of suppression: while the linear ramp-down reaches the suppressed state sooner, it relinquishes late in the episode, whereas the learned schedule holds it to the end. In the inverse, zonal-break task, the agent rediscovers the physically anticipated optimum: an interpretable up-down antisymmetric pattern $[-9,-9,+9,+9]$. This pattern imposes an electric field along the flux surface, driving a radial $E\times B$ convection that disrupts the zonal coherence and re-establishes cross-field transport. The disrupted turbulent state is sustained in all five initial conditions over the evaluation horizon. Moreover, an action-space sweep shows that the high-reward region of the zonal-break task is a thin antisymmetric manifold occupying a vanishing fraction of the action space. This restrictive geometry defeats naive random exploration and motivates the physics-informed warm-buffer initialization adopted here. Such geometric constraints are likely generic for turbulence-control problems whose optimal actuation exploits the $E\times B$ geometry of the underlying flow.


Through this study, these two tasks point to what we consider the defining difficulty of learning control for a turbulent system: the learning signal must be extracted from a chaotic state. Individual reward evaluations may carry inherent fluctuations comparable to the differences the agent must resolve, and the reward landscape away from the physics-selected manifold can be nearly flat, leaving little usable gradient to follow. These challenges are evident in the raw reward's susceptibility to amplitude inflation and the failure of random exploration to reach the optimal manifold. What makes the problem tractable is supplying physical insight where the learning signal is weakest: in the definition of the reward and the initialization of the search. We expect this requirement to be generic, which strongly argues for treating physics-informed, rather than model-agnostic, reinforcement learning as the practical route to turbulence control.

As an exploratory study, this work demonstrates the potential of RL as a direct optimizer for nonlinear plasma dynamics, acting on the turbulent transport itself rather than on a linear or gradient-based proxy. In the future, we plan to incorporate more realistic simulation environments (such as three-dimensional and geometrically realistic edge-turbulence models, and ultimately experimentally accessible actuators like biasing electrodes and localized RF heating), and to develop surrogate environments to further accelerate the control loop.

\begin{acknowledgments}
B. Zhu is supported by the U.S. Department of Energy (DOE) under Grant DE-SC0026529. This research used resources of the National Energy Research Scientific Computing Center, a U.S. DOE Office of Science User Facility under Contract No. DE-AC02-05CH11231 (FES-ERCAP 0036750).
\end{acknowledgments}

\section*{Data Availability Statement}
The data that support the findings of this study are available from the corresponding author upon reasonable request.

\appendix
\section{Solver Benchmark}\label{sec:appendix}
\label{sec:benchmark}
The jaxHW solver is verified against GDB~\cite{zhu2018gdb} on the original MHW system, i.e., Eqs.~\eqref{eq:gov} without zonal drag ($\gamma_\text{ZF}=0$) and actuation ($\phi_\text{ext}=0$). This comparison is performed at the two regimes studied in the main text ($\alpha=0.1$ and $0.5$), using the same numerical configuration ($256\times256$ mesh, $64\times64$ domain, $\kappa=1$, $\mu=2\times10^{-5}$ with $\nabla^6$ hyper-diffusion). To make the comparison deterministic and independent of random-seed conventions, all runs are initialized from the same analytic Gaussian wave packet centered in the domain,
\begin{equation}
\phi_0 = A\, e^{-r^2/2\sigma^2}\cos\!\big(k_0 (y-y_c)\big), \qquad
n_0=\phi_0, \qquad \varpi_0=\nabla^2\phi_0 ,
\label{eq:wavepacket}
\end{equation}
with $A=10^{-3}$, $\sigma=8$, $k_0=1$ (near the fastest-growing drift wave) and $\varpi_0$ evaluated in closed form. Each code is run with two distinct time integrators, yielding four setups per $\alpha$ value. Specifically, we configure GDB with its native trapezoidal leapfrog ($\Delta t=5\times10^{-3}$, implicit spectral hyper-diffusion) and with a classic fourth-order Runge--Kutta scheme ($\Delta t=2.5\times10^{-3}$, hyper-diffusion evaluated spectrally and treated explicitly). Conversely, we configure jaxHW with RK4 ($\Delta t=10^{-2}$) and a trapezoidal-leapfrog scheme ($\Delta t=5\times10^{-3}$), both utilizing finite-difference hyper-diffusion.

All diagnostics are computed from the raw $(n,\varpi,\phi)$ frames of every run using a single post-processing implementation with spectral derivatives, so no per-code convention bias the results. With $\langle\cdot\rangle$ denoting the domain average and $\bar\phi$, $\tilde\phi$ representing the zonal and non-zonal components defined in Sec.~\ref{sec:problem}, we evaluate
\begin{gather}
E_i=\tfrac12\langle n^2\rangle,\qquad
E_k=-\tfrac12\langle \phi\varpi\rangle,\qquad
E_{kz}=\tfrac12\big\langle (\partial_x\bar\phi)^2\big\rangle,
\label{eq:diag_energies}\\
\Gamma_n=-\kappa\langle n\,\partial_y\phi\rangle,\qquad
D_\alpha=\alpha\big\langle(\tilde n-\tilde\phi)^2\big\rangle,
\label{eq:diag_fluxes}\\
D_n=\mu\big\langle n\,(-\nabla^2)^3 n\big\rangle,\qquad
D_\varpi=-\mu\big\langle \phi\,(-\nabla^2)^3\varpi\big\rangle,
\label{eq:diag_diss}
\end{gather}
which satisfy the energy balance $d(E_i+E_k)/dt=\Gamma_n-D_\alpha-D_n-D_\varpi$.
\label{eq:dEdt}

\begin{figure}[!htb]
\centering
\includegraphics[width=0.9\textwidth]{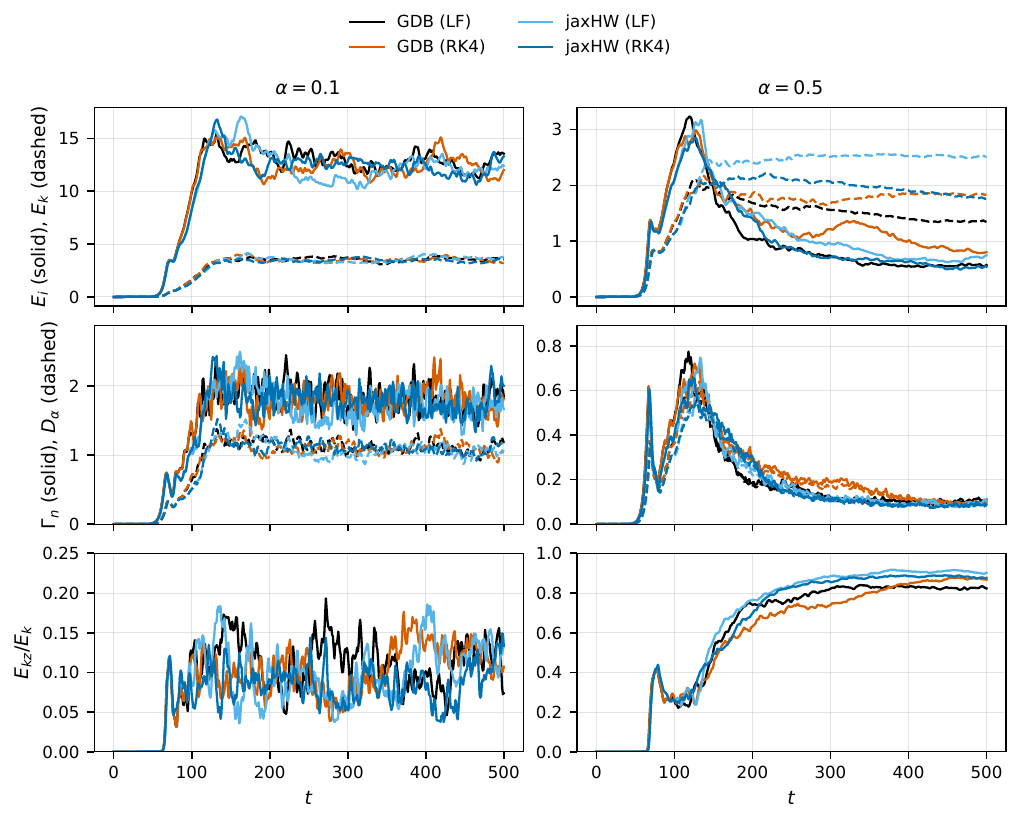}
\caption{Cross-code benchmark from the wave-packet initial condition, Eq.~\eqref{eq:wavepacket}: GDB and jaxHW, each with leapfrog and RK4 stepping, at $\alpha=0.1$ (left) and $0.5$ (right). Each panel overlays the four setups; colors distinguish the runs (GDB in black/orange, jaxHW in light/dark blue) and line style distinguishes the quantity (solid $E_i$, $\Gamma_n$; dashed $E_k$, $D_\alpha$). Bottom row: zonal-energy fraction $E_{kz}/E_k$.}
\label{fig:benchmark_mhw}
\end{figure}

Figure~\ref{fig:benchmark_mhw} summarizes the comparison. All four setups are indistinguishable through the linear growth phase. The two GDB integrators agree to a relative $L_2$ difference below $2\times10^{-5}$ over the growth phase ($t\le60$), confirming time-step convergence. The sharp transient near $t\approx68$, the first breakup of the coherent packet, is reproduced by all codes at the same time and amplitude: the peak flux agrees to four digits between the two GDB integrators and to within $2\%$ between jaxHW and GDB. Both jaxHW integrators track GDB to within $10-20\%$ through the growth phase, essentially the same margin: the discrepancy is largely due to the chaotic nature of turbulence. In the saturated phase all four setups agree statistically, with mean levels matching to within the realization scatter of the slow zonal modulation (single realizations per setup), and the zonal-energy fraction, the regime discriminator, reaches $E_{kz}/E_k\approx0.1$ at $\alpha=0.1$ and $\approx0.8$ at $\alpha=0.5$ in every code.

A primary motivation for the JAX-based solver, jaxHW, is that existing solvers are too slow for the many-query rollouts required by online RL training. 
For context, the Fortran 90 production code GDB~\cite{zhu2018gdb} advances the MHW model with a $256\times256$ mesh at approximately $4.16$~ms per step with the trapezoidal-leapfrog method on a Perlmutter CPU node, and the NumPy backend Python version of the same model takes about $16.95$~ms per step. However, on a Perlmutter A100 GPU node the JAX backend reaches $0.36$~ms per step with the leapfrog scheme and $0.62$~ms per step with RK4, roughly $12\times$ and $7\times$ faster than GDB, respectively. These are per-step timings at a fixed step count, reported to quantify the wall-clock cost of the RL training loop rather than as an absolute cross-code benchmark: the backends employ different time integrators and stable step sizes, so per-step ratios do not translate directly into per-physical-time throughput. The practical advantage of jaxHW for online RL lies in its low per-interaction cost and the seamless integration (i.e., the solver runs end-to-end on the GPU alongside the agent networks, eliminating host-device transfers, supporting JIT-compiled batched rollouts, and remaining differentiable for future model-based extensions).


\bibliography{ref}

\end{document}